\documentclass[twocolumn,aps,prl,showpacs,superscriptaddress]{revtex4}
\usepackage[dvips]{graphicx}
\usepackage{bm}

\begin{document}

\title{
Spin glass transitions of smectic-$A$ crosslinked elastomers }

\author{L. V. Elnikova}
\affiliation{
 A. I. Alikhanov Institute for Theoretical and Experimental Physics, \\
B. Cheremushkinskaya 25, Moscow 117218, Russia}

\date{\today}

\begin{abstract}
Elastomers are artificial polymeric materials created for industrial and commercial applications.
Depending on their purpose, they are performing in different species and structure modifications.
Our studies focus on the systems of elastomers randomly standing-distributed in a smectic $A$ (Sm$A$) liquid crystal.
Basing on the suggestion following from the experiment, that at a phase transition from Sm$A$ to nematic phase caused by an increase of a crosslink concentration, such a system survives a percolation transition at low crossilink concentrations, we propose a modeling explaining this phenomena.
We approve the three-dimensional Villain spin glass model and apply lattice Monte Carlo (MC) techniques on differential forms on a dual lattice, that is an alternative of a replica trick, developed for nematic elastomers in the 3D $XY$ universality.
In the results of that we have confirmed a concentration phase transition of percolation nature at a small crosslink concentration ($\sim$ 10 weight $\%$).
\end{abstract}

\maketitle

\section{Introduction}
A great number of quantum phase transitions in polymers and liquid crystals exhibits a percolation phenomenon.

Particularly, percolation properties of the transition from smectic-$A$ (Sm$A$) to nematic ($N$) phase in
crosslinked smectic elastomers were observed on the experiments on X-ray \cite{Ostr2004} and $^1$H-NMR
spectroscopy \cite{Sotta1}. Percolation in such systems may occur at the increase
of the crosslink density and/or at the increase of temperature as well.

Elastomers are polymer networks composed by crosslinked polymer
chains. They consist of nodes and links obeying ascertained rules of
self-organization \cite{deGennes_book, Khokhlov_book, Grosb_UFN}. Polymeric random network systems are differing with types of
ordering, orientational and translational, caused by topological
defects (dislocations, vortexes etc.) \cite{Ostr2004,
7, 8}.

For mere polymers, the percolation problem is often discussed in the light of the
analogy to dimers and connective conductors, proposed by de Gennes \cite{dG_76}, \cite{Kasteleyn}, \cite{Broderix_1997}, and
fractal dimensionality.
However, crosslinks are not interacting via only central forces (refs. in \cite{Sotta_ima}) (which correspond to a bending energy), but both strength and bending energies are presented, carrying out to a high percolation threshold. At any modifications in a polymeric network, using of the de Gennes's theory may become more embarrassing.

In mere smectics, the linear nontrivial disclinations (3D vortexes or
dislocation loops) as well as point monopole-type defects take place \cite{Polyakov, Leo, Dasgupta}.
From a representation of topological defects in lattice theories, percolation associates with condensation of vortexes in the high
temperature deconfinement phase \cite{Polikarp49}. Where, similarly that in polymers, the fractal-dimension domains of this phase will exist on a non-integer lattice, the Polyakov line plays the role of the order parameter.

For pure smectic-$A$ liquid crystals, moving by the second
order tilt transition to the nematic phase, Dasgupta proposed the loop inverted
analog of the superconductive $XY$ model \cite{Dasgupta}.
The action of this model is equivalent of the Villain's action
\cite{Villain}, if in the anisotropic $XY$ model, the de Gennes coefficient $K_1$ equals
to zero \cite{Dasgupta, Nelson1981, Dasgupta_perc}.
Dasgupta (\cite{Dasgupta} and references therein) derived mapping of the partition function of the lattice de Gennes model onto
that of a dislocation loop model. A vortex, as well as a dislocation with its similar properties, expresses a non-trivial classical
minimum of the action \cite{Polyakov} of a scalar field theory $\Phi$ with
broken global $U(1)$ symmetry, and the $XY$ model is its
quantized variant.

Relating to the 3D $XY$ universality, percolation transitions could be described via the $U(1)$ gauge field $A$ \cite{Polikarp}
on a dual cubic lattice at critical temperature $\beta_c \approx 0.4542$ in appointed fractal dimensionality $D_f$, while
quantitative characterization of the order parameter is remaining complicated by ambiguity in the field $A$ \cite{Dasgupta}, \cite{Nelsen}.

In a general case, for spin glass models of polymer networks, one applies the replica methods \cite{Derrida,
Parisi_Virasoro_book, Dotsenko_book, Nishimori_book}.

With understanding the Sm$A$--$N$ transition in
alkylcyanobiphenyl $n$CB (in particular, in 8CB) caused by porosity, it has become possible to subdivide the theory of Sm$A$--$N$ transitions into the three schemes (Anopore, silica Aerogels and Vycor glass) \cite{Leo, Qian}.
As was shown numerically, in bulk nematics \cite{Qian}, due to the coupling between smectic and nematic order parameters and director's fluctuations, the Sm$A$--$N$ transition deviates from 3D $XY$ universality to a nonuniversal crossover value between 3D $XY$ and a tricritical behavior with the critical exponent $\alpha$ of the effective heat capacity ($\alpha$ ranging between 0.26 and 0.31 confinements).

In the lack of elastomers, in the 3D superconductive universality \cite{Leo, sc_vort} with a scalar gauge field $k=0$, a
dimensionality of defects (vortices) is $D-k-2=1$
\cite{Polikarp}, and may be successfully assigned to aerogels with the smectic ordering. A further fractal contribution into the additional material (defectless) dimensionality may be produced by the crosslinked elastomers, confined in such smectics.

So the order parameters of both the smectic-$A$ and the polymer
network is to be conflated as a whole. In this paper, we present
the Sm$A$--$N$ transition in the system of
smectic$A$ and crosslinked elastomer as behaving in frames of the Villain model and specify the spin-glass treatment of it.

\section{The Villain model on a dual lattice}
Except the continual theory of the thermotropic Sm$A$-N transition in crosslinked elastomers \cite{Olmst_Terent}, the related thermodynamical expression in similar Sm$C$ systems was given in \cite{Ter2002SmC}.

Referring to the
continual model \cite{1997replica}, the authors of \cite{Ter2002SmC} have found,
that though in lack of the external magnetic field,
the expressions of the free energy density for the smectic and nematic elastomer are analogous, however, the nematic elastomers possess so called "soft elasticity", e. g. there is no homogeneity of a nematic director field, but splay, bend, and splay-twist fluctuations are presenting together. This fact, as well as the conclusions of \cite{Sotta_ima} from the "spring-ball" $ $ numerical modeling of gels and filled polymers, being crowned with a non-affinity in the deformations here, compels us to accept a proposition of \cite{1997replica},
that spin glass models are only satisfying to receive statistical values revealing the content dependent percolation threshold in the Sm$A$--crosslinked elastomer.

Then we will introduce an assignment of lattice spin variables to any components and configurations of this system.
On its Abelian discrete $Z_N$ group $\exp\{2ik\pi /n\}$, $k=0,\ldots, n-1$, an oriented link
bordering plaquettes with the appointed orientation and non-orientable surfaces are permitted for even $n$.

Vortices of the Villain model are decoupled with spin waves, but the $XY$ model has the same topological characteristics,
as the Villain model. It is reasonable to associate frustration effects with a crosslink configuration, i.e. with unfrustration networks, where the thermal fluctuation does not destroy the long range order, according to the spin glass topology in three and more dimensions \cite{Koma, Bovier}. As to the percolation phenomena in these elastomers at the $SmA$-$N$ transition, a random field replaces onto
a boundary effect on the unfrustration network.

Further, we follow these ideas on a frustration for spin glasses \cite{Koma, Bovier}: "a frustration network is dual to a set of the $(d-2)$-cells in the dual lattice which form $(d-2)$-dimensional complexes which are closed or whose boundaries end at the boundaries of the dual lattice $\Lambda^*$ to the lattice $\Lambda$."

From the expression
\begin{equation}\label{1}
\prod_{p\subset c}\phi \partial p=\prod_{\langle i,j\rangle\subset c}(\hat{J}_{i,j})^2=1,
\end{equation}
one follows that an even number of the plaquettes must be frustrated in a cube $c$, and the complexes of dual $(d-2)$-cells to the frustrated plaquettes can not end in any cube. Here in (\ref{1}), $\hat{J}_{i,j}=J_{i,j}/|J_{i,j}|$ mean the signs of the couplings $J_{i,j}$ in the Toulouse's introduction of frustration for a loop $c$ along the bonds of the lattice ($\phi(c)=\prod_{\langle i,j\rangle \subset c} \hat{J}_{i,j}$).

Percolation in such cases is formulated \cite{Koma} as a Bernoulli bond percolation
process on the $Z^3$ lattice with densities $x$ and ($1 - x$), respectively for the sorts of spins, the
bonds of negative couplings $J_{ij}$ percolate for $x$ near 1/2 (just like in $Z^2$). The thermal fluctuation does not destroy the long range order in the unfrustration network. Namely we can expect the existence of the spin glass phase on the $Z^3$ lattice for the density $x$ near 1/2.

In \cite{Koma}, there are considered the cube configurations whose two-cells are made of either some unfrustrated pairs of frustrated and some frustrated plaquettes, or all made of unfrustrated pairs, so that  the way of choosing unfrustrated pairs as two-cells is not unique, and in $Z^3$ the ground state is highly degenerate; one may only hold a condition that any loop which is made of the boundary bonds of the unfrustrated pairs should be homologous to zero in the cube.

Related studies for all frustrated plaquettes has been numerically provided in $Z^2$ for a large dice lattice ($L=$84) \cite{Fazio}.

The Hamiltonian of the $XY$ spin glass model on a cubic lattice is written
as follows:
\begin{equation} \label{fXY}
H_{fXY}= \sum_{\langle i,j\rangle}J_{ij}[1 - \cos(\theta_i - \theta_j - \mathbf{A}_{ij})].
\end{equation}
This model supports vortices \cite{Fertig} unbind at high temperature via the Berezinskii-Kosterlitz-Thoules (BKT) transition.
Here $\theta_i$ indicates the angle of the $XY$ spin at the $i$-the site and
the summation is taken over all nearest neighbor pairs, and
the coupling constants $J_{ij}$ obey such a distribution as the $\pm J$ or Gaussian function as in the Ising spin glass. We set the Boltzmann constant $k_B$ to unity.

On the lattice bonds, $A_{ij}\equiv -A_{ji}$, the magnetic currents connect the centers of lattice boxes and obey
\begin{equation}
A_{ij}=\frac{2\pi}{\phi_0}\int_{r_i}^{r_j} d\mathbf{r}\mathbf{A(r)}.
\end{equation}
These currents do not fluctuate, their sum belong the full lattice perimeter equals to $2\pi f$, $\sum_{kvadratic} A_{ij} = 2\pi f$,
 the equality $f=\frac{\phi}{\phi_0}$ expresses the
magnetic quanta $\phi_0$ and the magnetic current $\phi$ per
one lattice cell, in fact, $f\in [0,\frac{1}{2}]$.

The substitution $e^{C\cos\varphi_{ij}}\rightarrow
\sum_{-\infty}^{\infty}e^{-C(\varphi_{ij}-2\pi n)^2/2}$
(\cite{Fertig}, where $\varphi_{ij}=\theta_j-\theta_i$ is the rotation angle
of neighboring spins on a lattice, and $n$ is whole) made by Berezinskii and Villain (see references in \cite{Korsh_rev}, and \cite{Villain75}) allows us to write the spin glass partition function of (\ref{fXY}) on the original lattice:
\begin{equation}\label{BV}
Z = \int \mathcal{D}\theta
\sum^{\infty}_{p_{ij}=
-\infty}
\exp\{[-\frac{J_{ij}}{2}(\theta_\mathbf{j}-
\theta_\mathbf{i}+ 2\pi p_{ij})^2]\},
\end{equation}
here $p_{ij}$ are winding numbers lengthwise a loop, $\theta_\mathbf{i}$ and $\theta_\mathbf{j}$ mean phases in given lattice sites. $\mathcal{D}\theta$ denotes the integral over all link variables $\theta$'s.

Possessing view of the Coulomb gas partition function, (\ref{BV}) corresponds with the partition function derived in \cite{Dasgupta_perc} on a dual lattice describing a set of integrating vortex loops defined
in the generalized Villain's model with a non-zero "chemical potential".

With the technique of differential forms \cite{Polikarp49, Polikarp-Ivanenko} on the lattice, one builds the dual lattice
configurations corresponding to (\ref{BV}) with the next partition function
\begin{equation}\label{dual partition}
Z^{dual}= \sum _{k\in Z(C^0)}
\sum_{l\in Z(C^1)}\hat{J}\exp\{-\lambda'\|dk+l\|^2-\beta'\|l\|^2\},
\end{equation}
here the field of "currents" $ $ $l$ attached to the links and flowing along the $C^1$ circuit,
and the "monopole" $ $ field $k$ attached to the sites, which is provided with currents along $C^0$. The expression (\ref{dual partition}) also has an equivalent sense as $Z_{compact}=Z_{defects}\cdot Z_{noncompact}$.

Further connection between original and dual lattice variables are performed according to the transformation 
\cite{Polikarp49} in the gauge invariant form of $Z^3$ space. The terms $\hat{J}\exp\{-\lambda'\|dk+l\|^2-\beta'\|l\|^2\}$ reflect the $\exp\{-\frac{J_{ij}}{2}(\theta_\mathbf{j}-
\theta_\mathbf{i}+ 2\pi p_{ij})^2\}$ 's sense after the differential form transformation. Coulomb interactions between currents and monopoles
are straight to the linking number of currents \cite{Polikarp49}.

\begin{figure}
\includegraphics*[width=70mm]{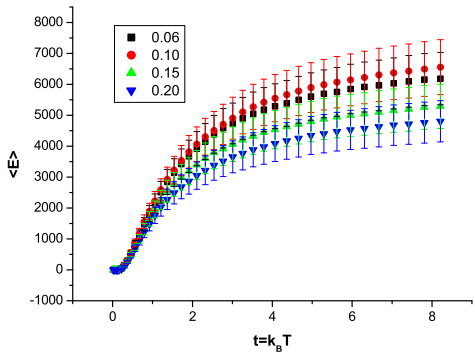}
\caption{\small Temperature dependence of an average energy at different crosslink concentrations}
\end{figure}

\begin{figure}
\includegraphics*[width=70mm]{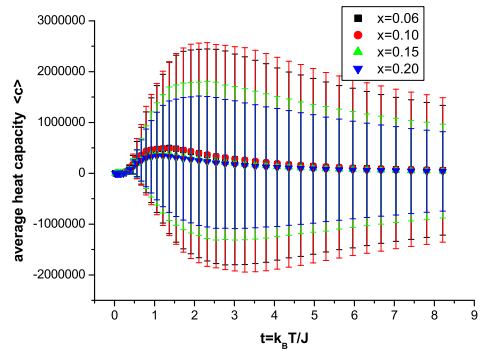}
\caption{\small Temperature dependence of an average heat capacity at different crosslink concentrations}
\end{figure}

\section{Numerical results and discussion}
In our MC modeling, we choose a lattice of size 48$^3$ and three type of whole spin variables, corresponding to smectic molecules (0), polymer networks (1) and crosslinks (-1), all occupied a cubic lattice sites. Closed Wilson loops and open loops appear in a system. Well-known standard loop algorithms were involved \cite{Heringa}, \cite{Stauffer}. The specific heat expression was improved in accordance to the Villain action properties \cite{Lang} (Figs 1-3 confirm, that temperature dependent transitions are not percolating).

For the four-dimensional $Z_N$ gauge theory, there was obtained the transformation connected the parameters $\beta'$, $\lambda'$ and $\mu'$ of the dual lattice with the original ones, which is follows \cite{Polikarp49}.

In the 3D case ($\mu'$, $\mu$ equal to zero), the reverse transformation for these coefficients of (5)
is performing by the equality $\beta=\frac{1}{4\lambda'}$ and vice versa.

\begin{figure}
\includegraphics*[width=70mm]{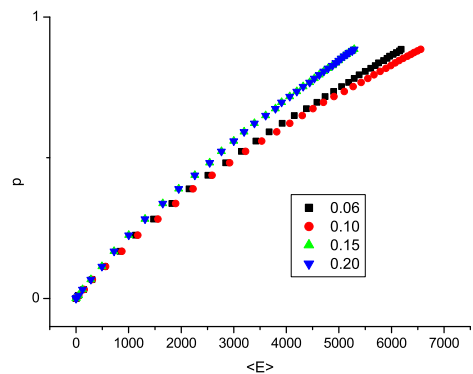}
\caption{\small Configuration dependence of probability from the MC algorithm on the dual lattice at different crosslink concentrations, 0.06, 0.10, 0.15, 0.20 $\%$}
\end{figure}

\begin{figure}
\includegraphics*[width=70mm]{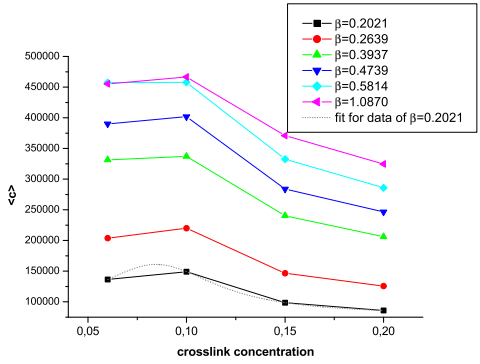}
\caption{\small Average heat capacity versus a crosslink concentration on the dual lattice at different temperatures in units $\beta=\frac{1}{T}$ (inset). Error bars are not shown. For visualization of a heat capacity extremums, a Lorentzian fit for $\beta=0.2021$ is shown}
\end{figure}

Fig. 4 illustrates extremums of an average heat capacity, which appear when a crosslink concentration surpasses
10 $\%$. This result confirms the experimental data of \cite{Ostr2004}, but does not satisfy a confirmation on percolation of such a transition: the configuration averages and their probabilities shows (Fig. 3), that they meet at temperatures close to zero, but not at $\beta=$0.4542 or 0.5, predicted in \cite{Polikarp} an \cite{Koma} in frames of the gauge BKT and spin glass theories.

It is known, that long-range ordering may be described in terms of spin glass models \cite{1997replica}, \cite{Nishimori} by means of the replica method, but, whereas our model is of nearest-neighbor interactions, we used a loop model. In \cite{1997replica}, there was predicted a phase transition in nematic elastomers via an order parameters, expressed as a magnetization, but in the version of a model as (\ref{dual partition}), this value does not exist.

\begin{figure}
\includegraphics*[width=70mm]{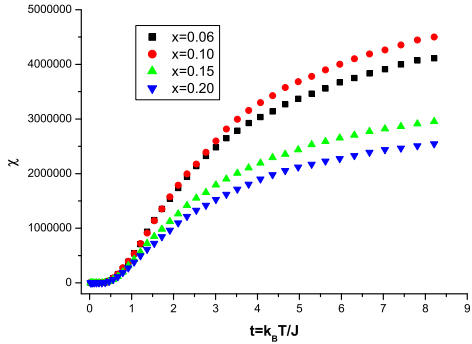}
\caption{\small Average susceptibility versus temperature on the dual lattice at different temperatures. Error bars are not shown.}
\end{figure}

\begin{figure}
\includegraphics*[width=70mm]{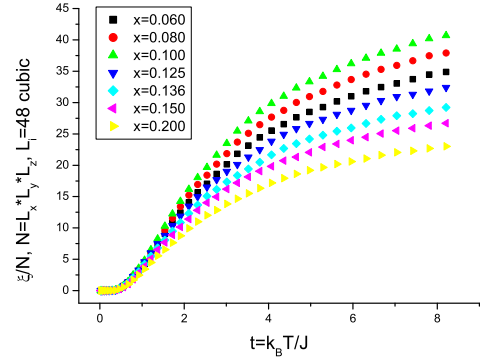}
\caption{\small The correlation function versus temperature on the dual lattice at different temperatures. Error bars are not shown.}
\end{figure}

\begin{figure}
\includegraphics*[width=70mm]{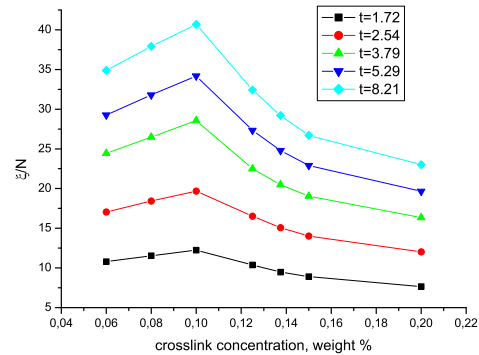}
\caption{\small The correlation function versus crosslink concentration on the dual lattice at different temperatures. Error bars are not shown.}
\end{figure}
Roughly speaking, using a monopole part of the Abelian Polyakov loop, we have got also correlators (Fig. 7) due to the standard procedure \cite{Boyko, Walter} as functions of a crosslink concentration, which do not directly confirm percolation, but manifest features increasing at growing temperatures at the 10-percent concentration of crosslinks.

\section{Conclusion}
To resolve the problem of percolation properties of elastomer configurations, we applied the technique lattice differential forms, builded on the Hamiltonian of the 3D $XY$ spin glass model, the main advantage of which is reduction of a dimensionality of the system and making an ability to estimate its thermodynamic values without cumbersome calculations. This method is alternative to the replica approach. However we have confirmed the experimentally observable percolation concentration phase transition at an increase of a crosslink concentration exceeding 10 percents.
$$
$$
\textit{Acknowledgements.} The author thanks Profs. B. I. Ostrovskii, M. I.
Polikarpov, M. N. Chernodub, and R. Savit for helpful discussions.

\end{document}